\documentclass[preprint,review,12pt]{elsarticle}

\usepackage{amsmath}
\usepackage{graphicx}
\usepackage{xcolor}

\journal{Journal of Materials Science and Engineering: C}

\begin{document}

\begin{frontmatter}

\title{Role of surface heterogeneous nucleation on nanoporous drug delivery systems}

\author[1]{Chiara Piotto}
\ead{chiara.piotto@unitn.it}
\author[1]{Paolo Bettotti\corref{cor1}}
\ead{paolo.bettotti@unitn.it}
\cortext[cor1]{Corresponding author}
\address[1]{Nanoscience Laboratory, Department of Physics, University of Trento\\
via Sommarive 14, 38123 Povo (Italy)}

\begin{abstract}
In this work we investigate the use of nanoporous carrier as drug delivery systems for hydrophobic molecules. By studying a model system made of porous silicon loaded with $\beta$-carotene, we unveil a fundamental limitation of these carriers that is due to heterogeneous nucleation that imposes a trade-off between the amount of drug loaded and the reproducibility of the release. Nonetheless, such issue is an alternative and improved method, compared with the standard induction time, to monitor the formation of heterogenously nucleated aggregates.
\end{abstract}


\end{frontmatter}

\section{Introduction}
Large amount of newly developed drugs are composed of poorly water  soluble molecules \cite{manikandanImprovingSolubilityBioavailability2013}. This fact causes a growing interest in developing drug delivery systems (DDS) able to handle poorly soluble compounds and, possibly, to protect them from degradation. Nanoporous inorganic carriers offers several advantages as they load large amount of drug and are (often) chemically inert. Yet, the release of hydrophobic molecules from such carriers has to cope with interfacial processes that can heavily impact their performance.
Here we investigate  a model system made by porous silicon (PSi) loaded with a highly lipophilic molecule ($\beta$-carotene, BCAR) and we discuss the general limitations induced by heterogeneous nucleation (HN) on release reproducibility.\\
BCAR has been chosen for its highly apolar nature (its solubility in water is 0.6 $mg/L$   at  25$^{\circ} C$), its easy oxidation -which permits to detect "subtle" chemical reactions taking place within PSi carriers- and its high optical absorbance -which allows its quantification down to $nmol$ concentration.\\
PSi possesses several attractive characteristics that render it a material of interest as DDS \cite{anglinPorousSiliconDrug2008,riikonenPorousSiliconDrug2015,liTailoringPorousSilicon2018}: its chemistry is well known \cite{leftwichChemicalManipulationMultifunctional2008}, it is biocompatible and biodegradable \cite{xiaPhotothermalBiodegradablePolyaniline2017} and it can be used as luminescent (tagged) carrier to track its fate once injected inside organisms \cite{daldossoOrangeBlueLuminescence2014}.\\
PSi might even protects delicate cargo, in fact: 1) its high absorbtion coefficient in the UV-vis range shields the drug from photochemical reactions; 2) as-prepared PSi surface easily undergoes  oxidization reactions, thus it can acts as sacrificial material to protect readily oxidized drugs;
3) the exchange of reactive gases (e.g. oxygen) is partially slowed down within the nanopores, thus a reduced amount of environmental active species is able to reach and degrade the loaded compounds.\\
Shielding properties of PSi are already demonstrated \cite{leBiogenicNanostructuredPorous2017,zuidemaOrientedNanofibrousPolymer2018,liHierarchicalStructuredProgrammed2018} for specific molecules, but we highlight that its supposed chemical stability needs to be carefully evaluated, since PSi might stimulate slow chemical reactions making this material unsuitable for drug storage.\\
Artificial surfaces with well controlled size and shape have been used to induce the HN of specific polymorphs \cite{reischlSurfaceInducedPolymorphismTool2015,diaoRoleNanoporeShape2011, stolyarovaModelEnhancedNucleation2006,diaoSurfaceDesignControlled2011}, but HN remains poorly investigated on nanoporous systems with random topography because of the difficulties in studying such irreproducible phenomena \cite{liuHeterogeneousNucleationHomogeneous2000}.\\
HN is commonly investigated using the induction method \cite{nanevTheoryNucleation2014} that poses formidable experimental issues in the case of HN on solid surfaces, as the definition of induction time depends on the spatial resolution achievable to check the presence of nuclei.\\
In this work we characterize fundamental interactions that take place between nanoporous carriers and hydrophobic molecules and we highlight the requirement of long term stability studies to prove the storage capability of inorganic, supposed inert, materials. Furthermore we underline a strong effect due to HN that induces the formation of slowly soluble aggregates, which, in turn, make the release kinetics irreproducible. This fact has to be considered during the design of porous carriers, otherwise it would rule out one of their main advantage that is their large loading capacity. Finally we propose such irreproducibility to be used as an alternative and more sensitive method, compared to the standard induction time, to characterize the formation of aggregates.

\section{Experimentals}
Silicon (100) wafers were supplied by University Wafers; all solvents were bought from Sigma-Aldrich: hydrofluoric acid (HF) is 48\% in water, tetrahydrofuran (THF) and toluene were of anhydrous grade, ethanol was ACS reagent grade. (3-Aminopropyl)triethoxysilane (APTES) were bought from Merck and used without further purification. BCAR were purchased from different vendors (Sigma-Aldrich, Alfa Aesar with purity $>$97\%) and, as soon as received, its oxidation state was always checked with optical absorption measurements to confirm the lack of detectable, already existing, oxidated species.\\
Two type of PSi samples were prepared using the standard electrochemical etching procedure: \textbf{1}) nSi-$0.02\ \Omega cm$ etched in $16\%$ HF, $16\%$ H$_2$O and $84 \%$ ethanol(v/v) at $82\ mA/cm^2$ current density and \textbf{2}) nSi-$0.01\ \Omega cm$ etched in $5.5\%$ HF in water added with $0.5\%$ TritonX-100 at $53\ mA/cm^2$ current density.\\
Four different PSi surfaces have been prepared:
\begin{description}
	\item[A:] as-prepared PSi: its surface is terminated with hydride species;
	\item[B:] PSi treated with mild thermal oxidization: PSi \textbf{A} was placed on hot plate at 200$^{\circ}$ for 4 h;
	\item[C:] silanized PSi: PSi \textbf{B} was immersed in APTES (1$\%$ (v/v) solution in toluene) and kept on an hot plate at 60$^{\circ}$ for 15 min. Then it was thoroughly rinsed with ethanol;
	\item[D:] wet oxidized PSi: PSi \textbf{B} was immersed in H$_2$O$_2$ (30$\%$ w/w) for 24 h.
\end{description}
The mild oxidation conditions used in \textbf{B} samples were chosen to keep the process compatible with highly nanoporous samples that own large specific surface but tend to collapse into bulk silica upon high temperature treatment. The further treatment (\textbf{C} and \textbf{D}) has been introduced to stabilize the PSi surface (see discussion below).\\
The loading procedure depends on the type of experiment. In the immersion method the drug was loaded by simply dipping the PSi sample for 24 h into THF BCAR solution. The impregnation of porous samples was done by subsequent release of some drops (few $\mu L$ each) of the loading solution, until the desired amount of BCAR was loaded onto the porous area, carefully controlling the spread of each drop and avoiding the wetting of the non porous region. Before depositing a new drop we waited for solvent evaporation.\\
Released BCAR has been quantified using a Varian Cary-100 spectrophotometer, using quartz cuvette and spectra were baseline corrected against the same cuvette filled with solvent.

\section{Results and Discussion} 
\begin{figure}
	\centering
	\includegraphics[width=0.7\columnwidth]{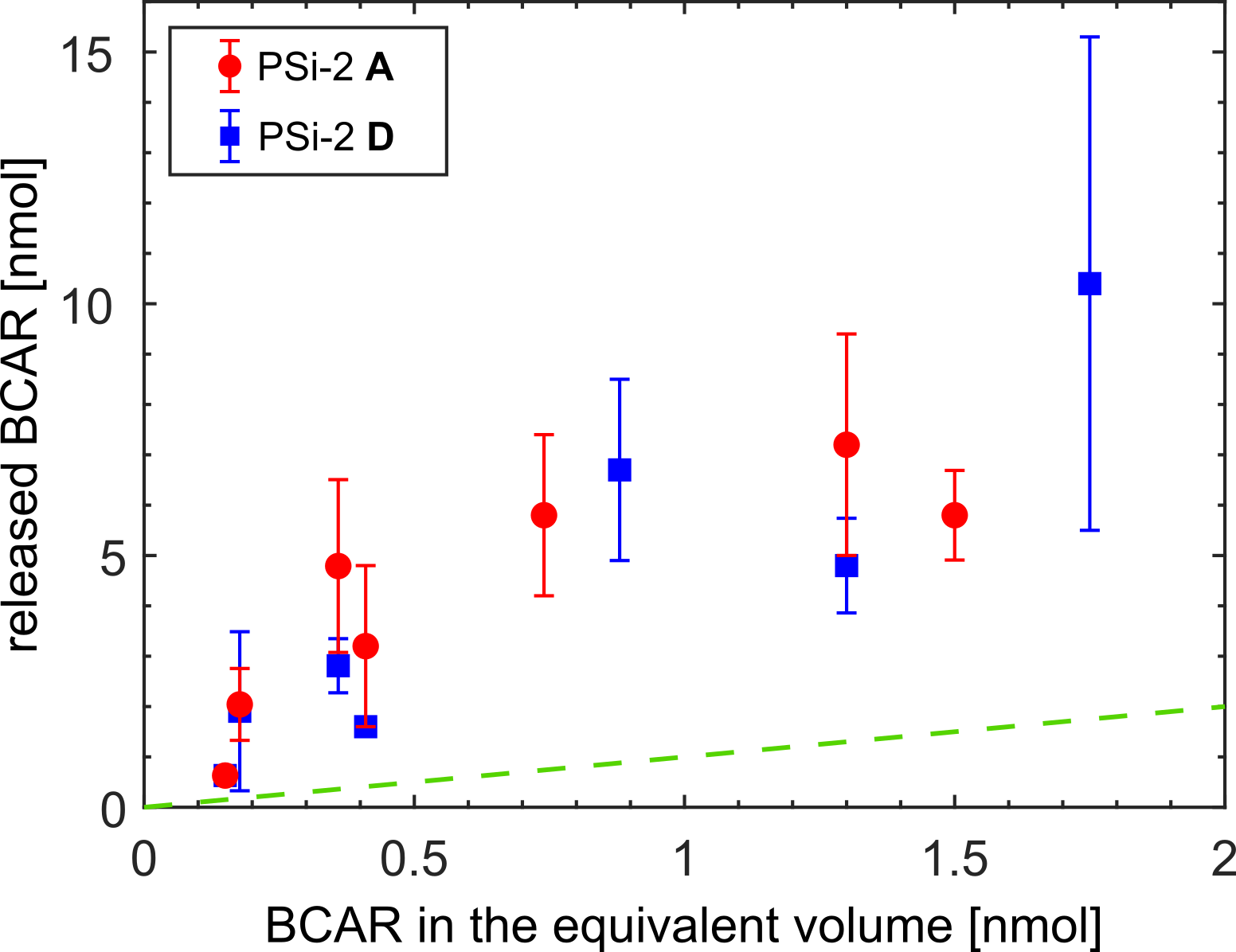}
	\caption{Quantity of released BCAR vs the estimated BCAR amount inside the pores at equilibrium. Green dotted line is the bisector.}\label{immersion}
\end{figure}

\begin{figure}[!t]
	\centering
	\includegraphics[width=0.3\columnwidth]{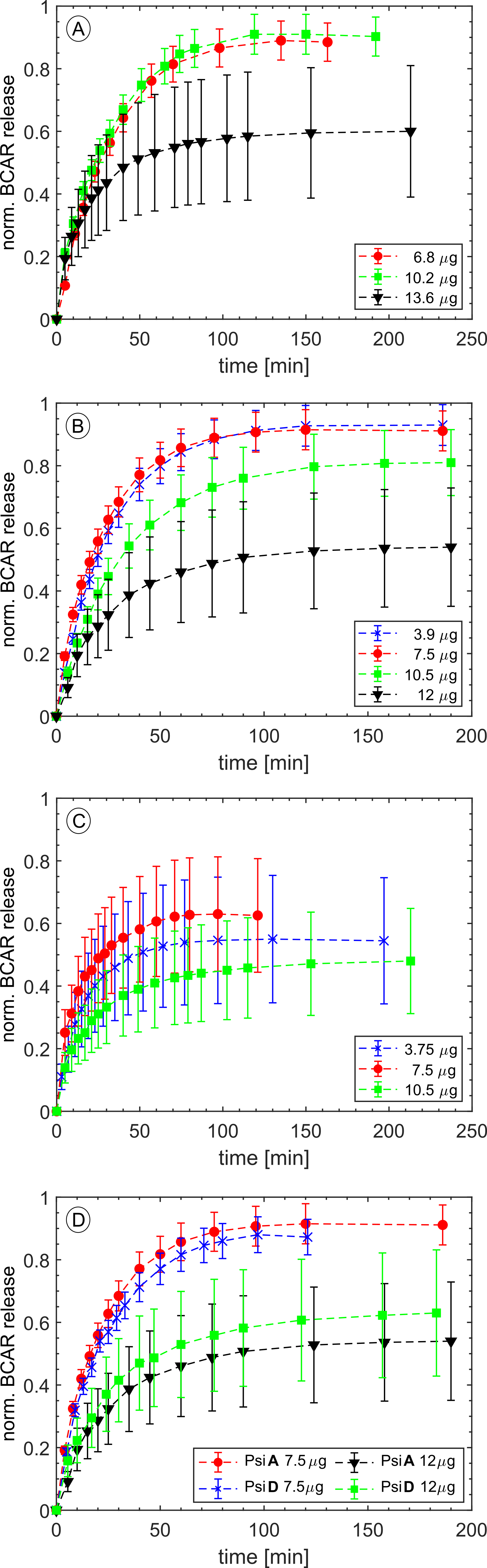}
	\caption{Normalised BCAR release profiles from PSi using loading solutions with BCAR concentrations of A) 0.17 $mg/mL$, B)  0.3 $mg/mL$ and C)  1.5 $mg/mL$. Panel D) reports the release from PSi-1 \textbf{A} and \textbf{D}. Lines are guides to the eyes.} \label{impregnation3}
\end{figure}

\begin{figure}
	\centering
	\includegraphics[width=0.7\columnwidth]{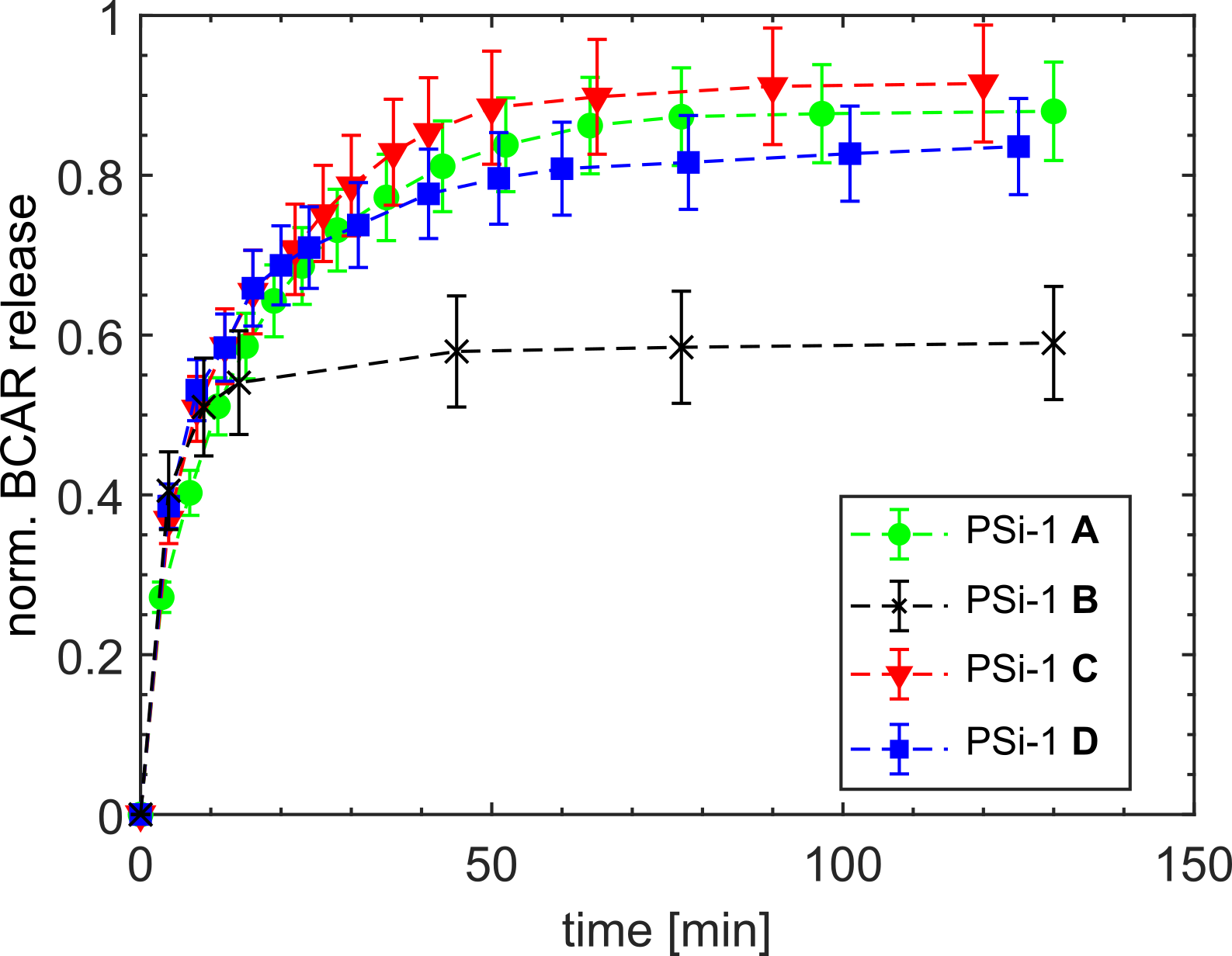}
	\caption{Normalised BCAR  release profiles from PSi-1 \textbf{A} (green-circles), PSi-1 \textbf{B} (black-crosses) PSi-1 \textbf{C} (red-triangles) and PSi-1 \textbf{D} (blue-squares); lines are guides to the eyes.} \label{impregnation2}
\end{figure}

Since the BCAR has a limited solubility in many solvents, we dissolved it in THF to prepare the initial, concentrated solutions used to load the samples. BCAR has been loaded into PSi with two methods: 1) immersion and 2) impregnation. We did not fragment the PSi layer into microparticles,
as it is usually done in DDS application,
since the use of flat samples eases the understanding of surface effects, such as aggregates formation.\\
Fig.\ref{immersion} reports the results obtained with the immersion method: PSi-2 \textbf{A} and \textbf{D} are immersed in THF solutions with different BCAR concentrations; then the non-porous surface of each chip has been rinsed with THF before placing the sample in THF. The amount of BCAR released is 5 to 20 times larger than the equilibrium concentration of an homogeneous solution without physical adsorption (indicated by the green dotted line, details of calculation are reported in the ESI). Neither the amount of BCAR entrapped, nor the SDs depend on the state of the PSi surface as they present the same behaviour both on as-prepared (hydrophobic) and on oxidized (hydrophilic) samples.\\
Since THF in an excellent solvent (BCAR solubility: 10 $mg/mL$), we assume that BCAR is released quantitatively: on average, roughly 5 nmol of BCAR have been loaded in each sample, i.e. about $10\ \mu g /mg_{PSi}$ that is an order of magnitude smaller than the equilibrium amount reached by loading hydrophilic drug into hydrophilic PSi  \cite{fryOxidationinducedTrappingDrugs2014}. The large SDs obtained (average value: 30\%) can not be ascribed to non-specific adsorption, otherwise similar SDs should have been obtained also from impregnation releases for samples   loaded with the same amount of BCAR (see data reported in Fig.\ref{impregnation3}B: PSi loaded with 3.8 $mg$ using 0.3 $mg/mL$ shows a SD of about 10\%). This comparison suggests that the high SDs of immersion experiments are due to aggregates that form inside the pores, even if BCAR is loaded at concentrations lower than its solubility in THF.
The large loading irreproducibility obtained using the immersion method does not allow for precise delivery of known amount of drugs, thus we investigated if impregnation method provides a better control.\\
We used THF solutions of BCAR at various concentrations to impregnate each PSi-1 \textbf{A} samples with different amount of drug. PSi samples compositions are summarized in Tab.\ref{tab1}.  
\begin{table}
	\caption{Loading conditions of PSi-1 \textbf{A} samples.} \label{tab1}
	\centering
	\begin{tabular}{|c|cccc|}
		\hline
		BCAR loading  & \multicolumn{4}{c|}{loaded BCAR}\\
		sol. conc. [$mg/mL$] & \multicolumn{4}{c|}{[$\mu  g$]} \\[1ex]
		\hline
		1.5 & \colorbox{yellow}{3.8} & \colorbox{yellow}{7.5} & \colorbox{yellow}{10.5} & \\
		0.3 &  {3.8} &  {7.5} & 10.5 & \colorbox{yellow}{12.0} \\
		0.17 & &  {6.8} & {10.2}  & \colorbox{yellow}{13.6} \\
		\hline
	\end{tabular}
\end{table}
During the loading, no macroaggregates form on PSi surface and the coffee-stain effect is limited. In fact, roughly 48\% of the BCAR was loaded in the 45\% central area of the sample (details are given in the ESI).
Releases were monitored using ethanol as model solvent (BCAR solubility is 0.03 $mg/mL$) as it slows down the kinetics, compared to THF and it permits to follow the kinetics with sufficient accuracy. \\
As reported in Fig.\ref{impregnation3}A-C, different kinetics were obtained depending  both on the amount of BCAR loaded and on the concentration of the loading solution. Using the less concentrated solution  (0.17 $mg/mL$, see Fig.\ref{impregnation3}A) the samples loaded with 6.8 and 10.2 $\mu g$ of BCAR show nearly complete and reproducible BCAR release (about 90$\pm$6\%), while the sample containing the largest amount of BCAR releases only about 60$\pm$20\%. By increasing the concentration of the loading solution from $0.17\ mg/mL$ to $0.3\ mg/mL$ (see Fig.\ref{impregnation3}B), a reduced recovery of BCAR occurs at smaller loading (together with a corresponding increase of the SDs). With the most concentrated solution ($1.5\ mg/mL$, see Fig.\ref{impregnation3}C) only limited and poorly reproducible releases of BCAR are achieved (55$\pm$18\% on average), irrespectively from the loaded amount.
Moreover, during the experiments done on the samples highlighted in Tab.\ref{tab1}, macroscopic surface aggregates form, even if the BCAR final concentrations were from 6 to 23 times lower than its solubility in ethanol and the release media were well stirred. The solubility of aggregates is slowed down compared to BCAR in molecular form; in fact they take up to 35 hours to completely dissolve. We have no evidence of either their crystalline or amorphous form, as the amount of material was too small to collect sufficient X-ray signals.\\  
The presence of aggregates and their role on the release kinetics is highlighted by defining the following Figure-of-Merit: $FOM=\frac{[\mu g] \ \text{loaded BCAR}}{\text{SD of release experiment}}$. Large $FOM$ indicates that BCAR release is quantitative and reproducible, while small $FOM$ points to partial and irreproducible recovery.
\begin{figure}
	\centering
	\includegraphics[width=0.7\columnwidth]{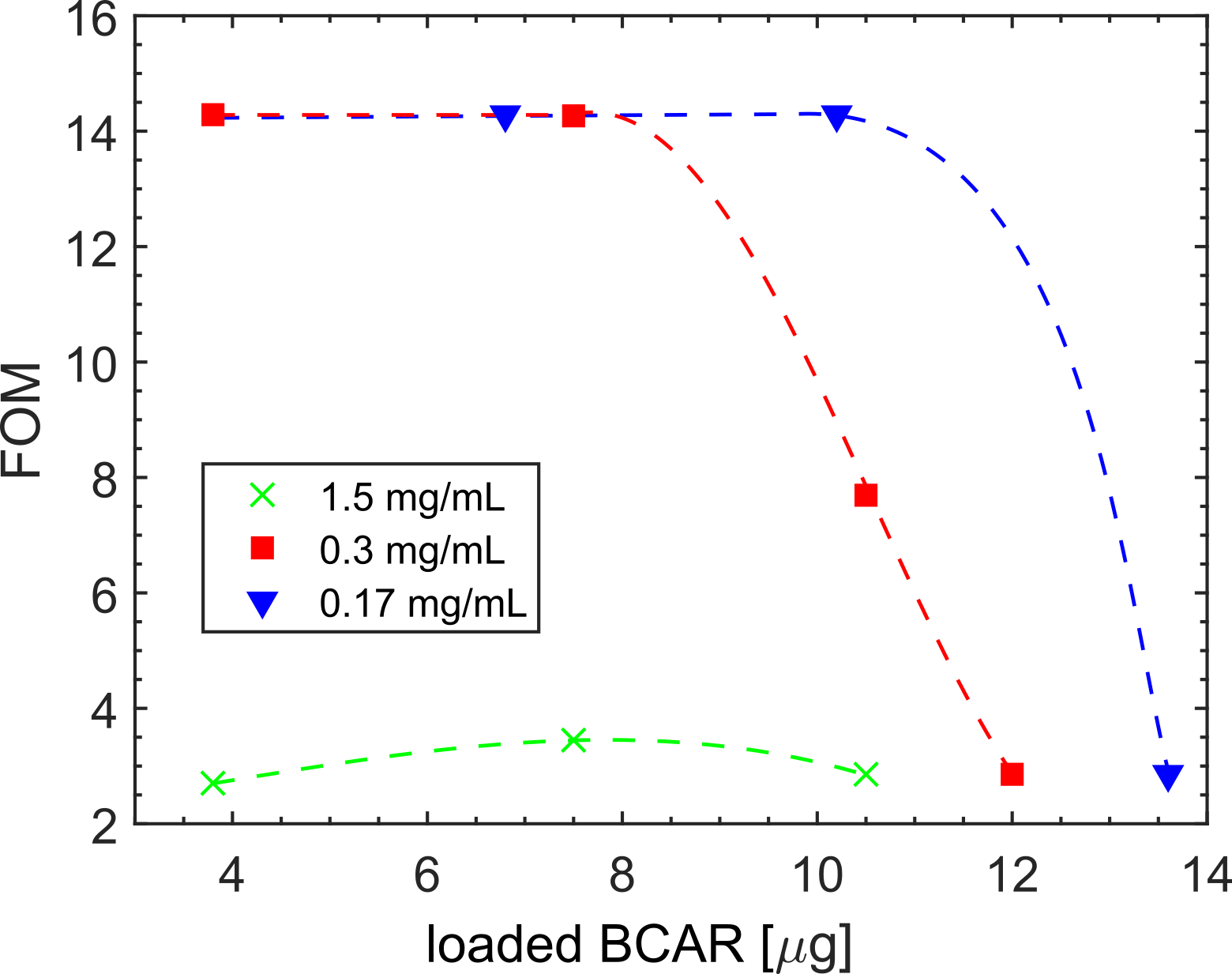}
	\caption{FOM as a function of the amount of loaded BCAR.} \label{fom}
\end{figure}
Fig.\ref{fom} reports how $FOM$ varies across all the dataset: when the 1.5 $mg/mL$ loading solution is used, it always assumes low values, while using the more diluted solutions, the FOM value is low only for large amount of BCAR loaded and it increases for smaller quantities.\\
This fact correlates with the macroaggregates formation on PSi surface: they are clearly visible when only 3.8 $\mu g$ of BCAR are loaded using 1.5 $mg/mL$, while if 0.3 $mg/mL$ is used, their formation is evident when at least 12 $\mu g$ of BCAR are loaded. A particular case is the sample loaded with 10.5 $\mu g$ of BCAR using the 0.3 $mg/mL$ solution: its FOM has an intermediate value of 8, the sample does not form visible aggregates, but it releases slightly less BCAR, compared to the samples loaded with smaller BCAR amount (about 80\% vs 90\%) and its SD is a bit larger (11\% vs 6\%). These considerations suggest that, when using a 0.3 $mg/mL$ solution, the  limiting amount of BCAR that can be loaded without macroaggregates formation is 10.5 $\mu g$.\\
We did not observed significant difference when either PSi-1  \textbf{A} or \textbf{D} are used (as reported in Fig.\ref{impregnation3}D), thus our experiments suggest that the BCAR nucleation-aggregation phenomenon is mainly ruled by the high PSi surface energy due to nanostructuring, while it is weakly related to the actual surface chemistry. Because of their stochastic length-scales, nanostructured surfaces form nuclei with broad size distribution: the smallest being readily re-dissolved, while the largest will live for significant time; in case of highly insoluble drug, they live much longer than the typical rate of release. In turn, such broad distribution reflects into broader range of kinetics, that are caught by an increased SD during the experiments. We speculate that, during the release, localized volumes at pore openings become sovrasaturated in BCAR and that such concentration fluctuations are the driving force to induce the formation of nuclei  (crystallization driven by such effect have been already demonstrated for proteins \cite{fermaniHeterogeneousCrystallizationProteins2013}).\\
The role the surface chemistry plays in BCAR HN has been investigate by performing the experiments on the four different PSi surfaces. We found that \textbf{A},\textbf{C} and \textbf{D}  samples loaded with 6 $\mu g$ of BCAR (using 0.3 $mg/mL$ solution) release almost quantitatively the drug with SDs of about 6\% (see Fig.\ref{impregnation2}); on the other hand \textbf{B} samples release only 59$\pm$7\% of the loaded molecules.
To understand the fate of the missing BCAR, after the release in ethanol, the \textbf{B} samples have been placed in THF, so that, if BCAR aggregates formed within the pores, they would  be readily dissolved. The facts that no BCAR has been detected in THF and that the SDs are less than 10\% suggest that the smaller recovery from \textbf{B} is not due to nucleation-aggregation phenomena. Since \textbf{C} and \textbf{D} samples released nearly completely the loaded BCAR, we believe that low temperature oxide alone is not a stable surface and it is able to degrade large amount of BCAR within short time (few hours). This fact is quite surprising, as \textbf{A} samples are generally thought to be more reactive than \textbf{B} ones. A simple immersion of the low temperature samples in peroxides or the APTES silanization is sufficient to stabilize PSi, albeit on the short time scale of load-and-release experiments. In fact, the BCAR long term stability is limited in all the PSi samples: we impregnated PSi-1 \textbf{A} and \textbf{C} and we performed the release in THF after two weeks, during which the samples were stored in the dark and at room temperature; the amount of BCAR recovered were 15$\pm$4\% and 40$\pm$7\%, respectively. The low BCAR stability  might be the result of autooxidation reactions, as reported in \cite{burton2014beta} and we speculate that reactive species (possibly radicals) and/or defects present on the surface of the low temperature oxides allow tunneling of electrons from BCAR to the semiconductor, speeding up the degradation of the drug.

\section{Conclusions}
By discussing the role of HN on nanoporous materials used as DDS for hydrophobic molecules, we underline two main limits of these class of DDS.\\
The first point to be carefully controlled is the capability of the carrier to effectively store and protect the compounds over reasonably long time scale: the simple load-and-release experiment does not show the (possible strong) cargo degradation induced by slow chemical reactions.\\
The second limit is the negligible amount of drug that can be loaded (irrespective from the state of the surface) without the formation of macroscopic surface aggregates during the release in \emph{not-so-good} solvents (e.g. in biological relevant environments). The presence of aggregates was detected by a reduced amount of BCAR released and by a corresponding SDs increase. Such aggregates form even if drug concentration is well below the solubility limit and we suppose that nucleation events happens nearby pore openings due to localized sovrasaturated BCAR regions. The dissolution rate of these aggregates is highly variable and much slower than the rate at which molecular BCAR is released from PSi under same conditions. The poorly controlled aggregation degrades the reproducibility of the experiments, thus the delivery of hydrophobic molecules from nanoporous carriers needs to find a the trade-off between the amount of loaded drug and the release reproducibility.\\ 
We suggest the use of such experiments to check for HN events on nanostructured surfaces as an alternative to the standard induction time method. Our approach shows several important advantages as it does not rely on visual inspection of the samples and is  able to detect solid nuclei well before they reach the -generally considered- micron-size regime. Compared to the induction time, this method provides an average response over the entire sample surface on a much shorter time and pave the way to more in depth studies about surface driven HN. 

\section{Supplementary Information}
\subsection{Amount of $\beta$-carotene loaded with the immersion method}
To calculate the amount of $\beta$-carotene (BCAR) loaded into porous silicon (PSi) samples at equilibrium, both the BCAR concentration in the loading solutions and the empty volume of the Psi samples should be known.\\
The samples were loaded by submerging them in 1.5 $mL$ of BCAR solutions in tetrahydrofuran (THF) at known concentrations for 24 h (0.32, 0.38, 0.77, 0.88, 1.60, 2.80 and 3.22 $mg/mL$ for PSi-2 \textbf{A}; 0.32, 0.38, 0.77, 0.88, 1.89, 2.80 and 3.76 $mg/mL$ for PSi-2 \textbf{D}.\\
The empty volume in each Psi chip is about 25 $\mu L$, as estimated from the porous area ($A$=0.28 $cm^2$), the layer thickness and its porosity.\\ 
The thickness ($d$=16 $\mu  m$) and the refractive index ($n$=2) of PSi-2 samples were obtained from SEM cross section and optical interferometry.  Once these two parameters are known,  the layer porosity (55\%) was estimated via Bruggeman approximation:
\begin{equation*}
\label{Bruggemanformula}
f \cdot  \frac{n^2_{b}-n^2}{n^2_{b}+2n^2}+ (1-f) \cdot  \frac{n^2_{a}-n^2}{n^2_{a}+2n^2}=0
\end{equation*}
where $f$ indicates the fraction of silicon in the porous layer, $n_a$ and $n_b$ the air and silicon refractive index (in the IR region  $n_a$=1 and $n_b$=3.4).\\

\subsection{Coffee-stain effect in the impregnation method}
\begin{figure}
	\centering
	\includegraphics[width=0.5\columnwidth]{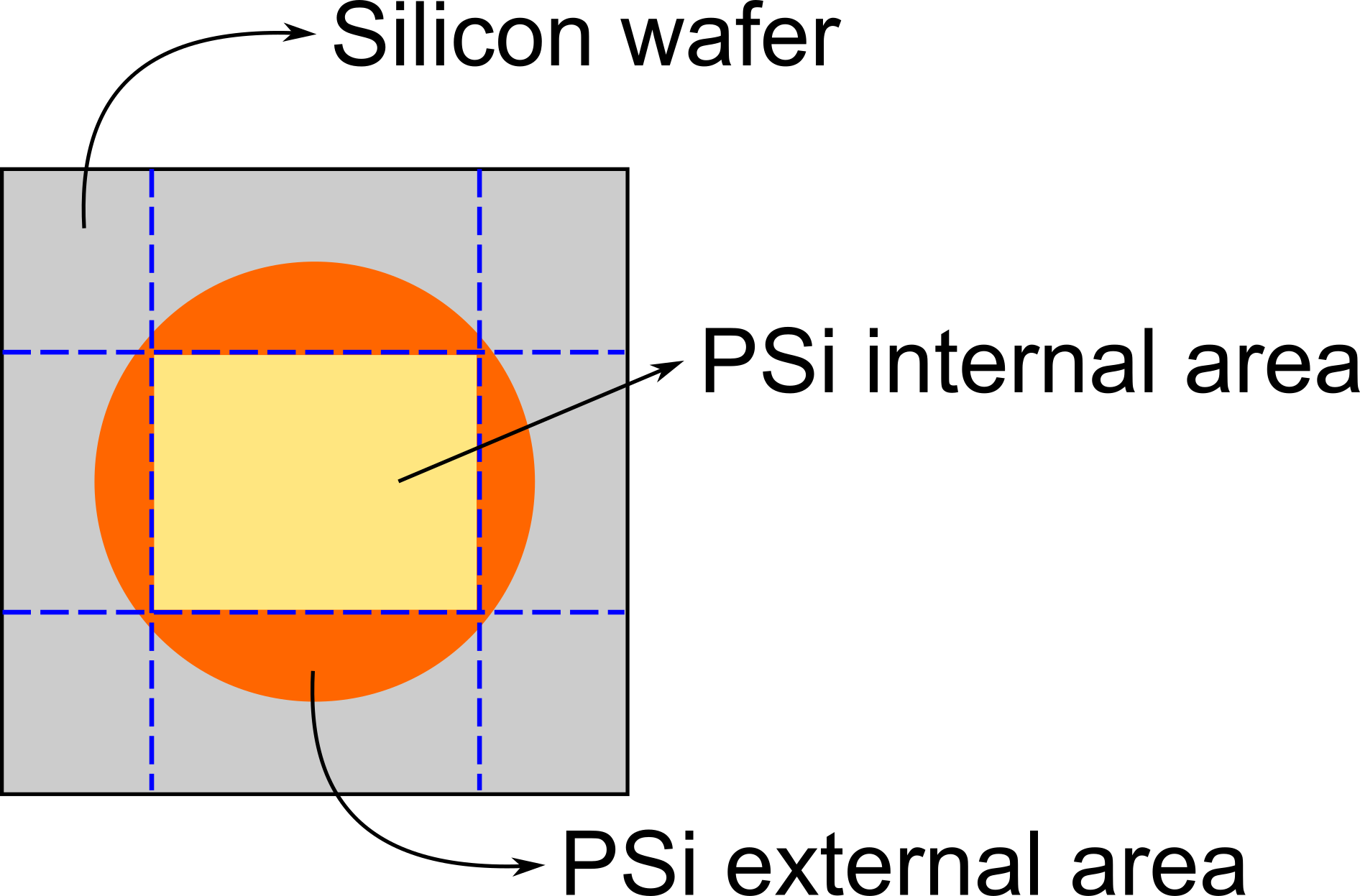}
	\caption{Schematic of PSi internal and external areas.} \label{disegno}
\end{figure}

Since the impregnation method requires the sequential deposition of some liquid drops on the PSi surface, we might expect the formation of coffe-stain on the edge of the porous area. To evaluate it, we loaded a PSi-1 \textbf{A} sample and we cleaved it along the blue dotted lines sketched in Fig.\ref{disegno}. Then we performed the BCAR release in THF both from the external part (55\% of the total area, orange area in Fig.\ref{disegno}) and from the internal one (45\% of the total area, yellow area in Fig.\ref{disegno}). We estimated a negligible effect of the coffee-stain since 52\% of the BCAR was released from the external area and 48\% from the internal one.

\end{document}